\documentclass[aps,prl,twocolumn,showpacs]{revtex4}

\usepackage{graphicx}
\usepackage{hyperref}
\usepackage{amsmath}
\usepackage{amssymb}
\usepackage{epstopdf}
\usepackage{bbm}
\usepackage{graphicx}

\newcommand{\mean}[2] {  \langle  #1 \rangle _{#2} }
\newcommand{\ket}[1]{  { | #1 \rangle}  }
\newcommand{\abs}[1] {  | #1 |  }

\begin{document}

\title{Experimentally accessing the optimal temporal mode \\of traveling quantum light states }

\author{Olivier Morin, Claude Fabre, and Julien Laurat}
\email{julien.laurat@upmc.fr}
\affiliation{
Laboratoire Kastler Brossel, Universit\'{e}
Pierre et Marie Curie, Ecole Normale Sup\'{e}rieure, CNRS, 4 place
Jussieu, 75252 Paris Cedex 05, France}

\date{\today}

\begin{abstract}
The characterization or subsequent use of propagating optical quantum state requires the knowledge of its precise temporal mode. Defining this mode structure very often relies on a detailed a priori knowledge of the used resources, when available, and can additionally call for an involved theoretical modeling. In contrast, here we report on a practical method enabling to infer the optimal temporal mode directly from experimental data acquired via homodyne detection, without any assumptions on the state. The approach is based on a multimode analysis using eigenfunction expansion of the autocorrelation function. This capability is illustrated by experimental data from Fock states and coherent state superposition preparation.\
\end{abstract}

 
\maketitle

Non-classical photonic states are the key resources for developing a variety of protocols, ranging from quantum communication and computing to quantum metrology \cite{illuminati06,obrien09}. In these protocols, a precise knowledge of the modal structure is essential as it strongly influences the success of the targeted operations. Indeed, the detection or the processing modes have to be precisely adapted to the resource.  For instance, homodyne detection \cite{Leonhardt}, a widely used tool for quantum state tomography, projects the impinging state into the mode of a so-called local oscillator, which has thus to be perfectly matched. Any mode mismatch will translate to losses. Similarly, in quantum information processing schemes, such as linear optical computing where light states are combined in optical circuits, the modal structure plays a central role \cite{Rohde05}.

Precisely controlling this modal structure has been a long quest for quantum optics. In some experiments, the temporal mode profile can be easily inferred from the setup features. In pulsed parametric down conversion for instance, the temporal mode is defined by the pulse shape. When using a continuous-wave optical parametric oscillator, the cavity bandwidth leads to a double-decaying exponential profile \cite{Nielsen2007_1}. However, even in these simple cases, the theoretical mode is only approximate given the imperfections and complexity of the setups. More critically, the pulse shape can also be strongly altered after its generation due to some propagation effects or additional frequency filtering may also change the optimal profile, leading to large mismatch between the expected mode and the actual one. In some experiments, the mode in which the light is emitted can even be harder to predict. This is the case for instance for photonic states generated from atomic systems, including from large atomic ensembles \cite{felinto}. 
 
Various techniques have been developed to infer the temporal wavepacket of single-photon states. One technique can rely on the use of photon counting to access the temporal statistics \cite{Hockel}. However, this straightforward approach is not always easy to implement due to low count rates and can be constrained by the limited photon-number resolution of available detectors for characterizing states involving higher photon-number contributions. Furthermore, it does not give access to any sign information in the modal structure. Recently, another technique based on homodyne detection, which enables a full tomographic reconstruction, has been proposed for ultrashort single-photon pulses. It relies on an adaptive scheme to iteratively map the mode into the one of the local oscillator using pulse shaping techniques \cite{Bellini2012}.  

In this letter, we investigate a practical and direct method to experimentally infer the temporal mode profile of traveling quantum light states, not only single-photon states, without prior information and without any optimization procedure. The method is based on homodyne measurements with a continuous-wave local oscillator, which enable to realize a multimode expansion via the autocorrelation function. The optimal temporal mode, or the existence of various independent modes, can then be inferred. We implemented this approach for different quantum state engineering experiments and report here data from the preparation of Fock states and coherent state superposition (CSS). Such a technique enables high-fidelity quantum state preparation and provides a variety of information about the multimode structure of a given state, as highlighted later by the case of a two-photon Fock state generation. 
 
 \begin{figure}[t!]
\centerline{\includegraphics[width=0.9\columnwidth]{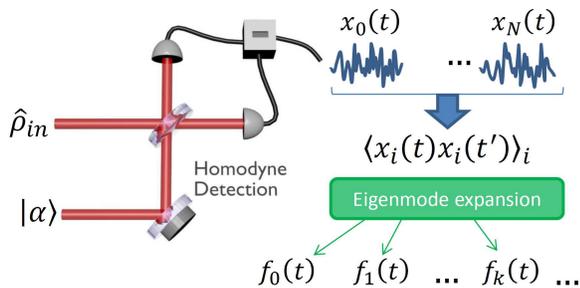}}
\caption{(color online). Schematic sketch. The propagating quantum state $\hat{\rho}_{in}$ is measured by homodyne detection with a continuous-wave local oscillator. The reported method consists in accessing the optimal temporal mode via a multimode analysis using eigenfunction expansion of the autocorrelation function  $\mean{x(t)x(t')}{}$.}
\label{fig1}
\end{figure}

To introduce the method, we first consider the case of a light field containing a single-photon state in a well-defined spatial mode.  It can be shown that single-photon states can always be considered as single mode states, in a mode that has to be determined \cite{Delaubert}. The question is to determine the temporal mode occupied by the single-photon by directly using the raw homodyne data. For each realization of the experiment, the homodyne detection, as illustrated in Fig. \ref{fig1}, provides a continuous measurement $x(t)$. This homodyne signal can then be processed with a temporal mode $f(t)$ to give a single quadrature outcome $x_f=\int f(t) x(t) dt$. One possible strategy to infer the optimal mode, as theoretically presented in \cite{Nielsen2007_1}, is to maximize the fidelity of the reconstructed state with a single-photon. However, this is not an easy task as it requires the use of some optimization algorithms in addition to the state reconstruction algorithm. A more straightforward method can be to maximize the variance of the measured quadrature in a given time interval. Indeed, the variance of the vacuum is $\mean{\hat{x}^2}{\ket{0}}=\sigma_0^2$ while it is three times larger for the single-photon state, $\mean{\hat{x}^2}{\ket{1}}=3\sigma_0^2$. The variance $\mean{\hat{x}_f^2}{}$ is thus a good parameter to maximize in order to infer the optimal temporal mode function. 

Interestingly, it can be shown that the variance $\mean{\hat{x}_f^2}{}$ of the filtered mode and the temporal mode $f(t)$ are linked to the autocorrelation function of the unfiltered homodyne signal by the expression:
\begin{equation}
\mean{\hat{x}_f^2}{}=\iint dt dt' f(t)f(t') \mean{\hat{x}(t)\hat{x}(t')}{}.
\end{equation}
The variance is thus expressed as a quadratic integral form. The kernel, given here by the autocorrelation function, $K(t,t')=\mean{\hat{x}(t)\hat{x}(t')}{}$, is symmetric and positive definite. As a result of the Mercer's theorem, it can be expanded in a series \cite{Courant}:
\begin{equation}
K(t,t')=\sum_{k=0}^\infty\kappa_k f_k(t) f_k(t'),
\end{equation}
where $f_k(t)$ are orthonormal eigenfunctions satisfying the completeness relation $\sum_{k=0}^\infty f_k(t) f_k(t')=\delta(t-t')$ and the eigenvalue equation
\begin{equation}
\int f_k(t')K(t,t')dt'=\kappa_k f_k(t).
\end{equation}
Any temporal mode $f(t)$  can be expressed as a linear combination of the eigenmodes, i.e. $f(t)=\sum_{k=0}^\infty\lambda_k f_k(t)$ with $\sum_{k=0}^\infty\lambda_k^2=1$. The quadrature operator of each mode is then defined by
\begin{equation}
\hat{x}_{f_k}=\int   f_k(t) \hat{x}(t) dt.
\end{equation}
Given this expansion, the variance for a given temporal mode $f(t)$ can be written as
\begin{equation}
\mean{\hat{x}_f^2}{}=\sum_{k=0}^\infty \lambda_k^2\kappa_k\leqslant \max_{k\in\mathbbm{N}}{\kappa_k}.
\end{equation}
It results from this expression that the maximum of variance is obtained for the eigenmode with the largest eigenvalue.

This multimode expansion has actually a more general physical meaning. As $\mean{\hat{x}_{f_k}\hat{x}_{f_{k'}}}{}=0$ for $k\neq k'$, it corresponds to an expansion over non-correlated modes. Choosing a temporal mode indeed realizes a single-mode measurement: the state is thus traced over all the other modes and the process would result in a statistical mixture if there were some correlations with other modes. The method introduced here is thus general and not only restricted to the single-photon case presented above. It can be applied to various quantum state engineering experiment for which extracting the optimal temporal mode is a central issue. Nevertheless, it is worth noting that we only consider here in phase correlation and the uniqueness of the expansion is only true if the eigenvalues are different. When eigenvalues are equal, but different from the one of the vacuum, this expansion does not guarantee that all the modes are separable and results in a \emph{partial} Schmidt-like decomposition. 

\begin{figure}[t!]
\centerline{\includegraphics[width=0.95\columnwidth]{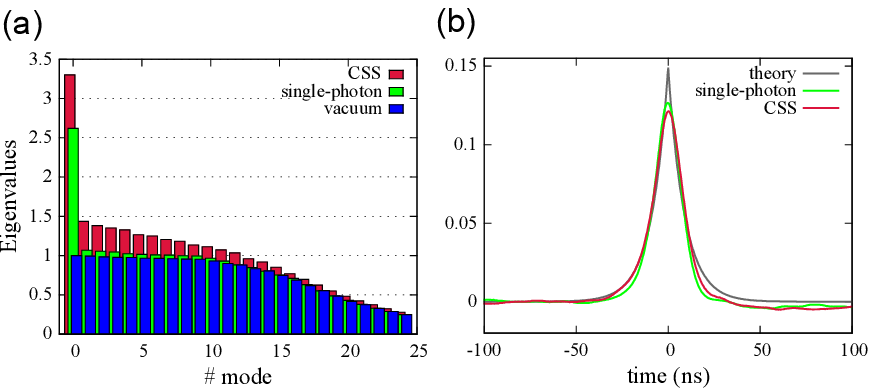}}
\centerline{\includegraphics[width=0.95\columnwidth]{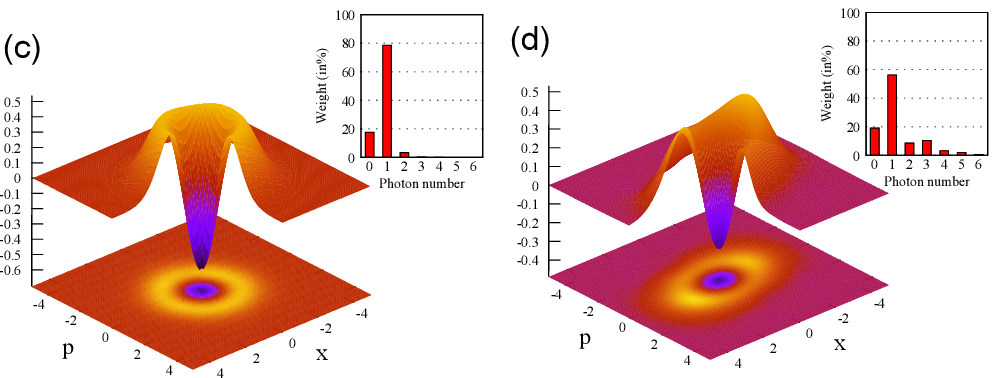}}
\caption{(color online). Single-photon and coherent state superposition generation. (a) Eigenvalues of the autocorrelation function $\mean{x(t)x(t')}{}$ (calculated from 50000 heralding events) given in blue for a vacuum state, in green for a single-photon and in red for CSS. Due to the finite bandwidth of the detection, the eigenvalues decrease for the modes with higher frequency components. (b) First eigenmode for single-photon in green and CSS in red. The black line gives the theoretical temporal mode corresponding to a double-decaying exponential profile. (c) and (d) Associated Wigner functions using the optimal temporal modes. The insets show the diagonal elements of the density matrices, without correction for detection losses.}
\label{fig2}
\end{figure}

As a first example, we illustrate this approach with the heralded preparation of single-photon from a two-mode squeezed vacuum emitted by a continuous-wave frequency degenerate type-II optical parametric oscillator (OPO) (cf. Ref. \cite{Morin12,Morin13}). The detection of a single-photon in one beam heralds the generation of a single-photon in the other one \cite{Hong86,EPJD,PRL}. Signal and idler fields are separated in polarization at the output of the OPO. The idler is frequency-filtered in order to remove non-degenerate modes due to the OPO cavity and then detected by a superconducting single-photon detector (SSPD, Scontel). Due to the continuous-wave nature of the pump, the temporal mode in which the single-photon is generated has to be determined. 

Experimentally, for each heralding event, the homodyne signal is recorded during 200 ns. The auto-correlation function is then computed from the recorded segments. This measurement being sampled (5 Gs/s rate), the experimental autocorrelation function is thus a real and symmetric matrix. The eigenfunctions and their corresponding eigenvalues are then computed numerically. Figure \ref{fig2}(a) provides such eigenvalues for the vacuum state and for the heralded single-photon. Let us first note that the eigenvalues for the vacuum are not all equal. This decrease for the modes with higher frequency components results from the finite bandwidth of the homodyne detection. Secondly, for the single-photon state, only the first eigenvalue is largely above the values of the vacuum state, as expected theoretically for such a state \cite{Delaubert}. The associated eigenfunction provides thereby the optimal temporal mode, as plotted in Fig. \ref{fig2}(b). This mode is then used for the state reconstruction \cite{maxlik} and the Wigner function corresponding to the heralded state is displayed in Fig. \ref{fig2}(c), together with the diagonal elements of the density matrix. We note that no correction for detection losses has been applied to these results. The single-photon component reaches here $79\pm1\%$, confirming the very high fidelity of the heralded state.

In order to compare to a model, we derive now the theoretical expression of the optimal temporal mode using the method based on the autocorrelation function expansion. In a type-II OPO, the signal and idler photons are emitted pairwise, leading to quantum correlations. Using the correlation functions for the annihilation and creation operators for the two modes, well-known for an OPO \cite{Reid,Nielsen2007_1}, one can derive the autocorrelation function of the ideal homodyne signal:
\begin{multline}
\label{eq1}
\mean{\hat{x}(t)\hat{x}(t')}{}=
\delta(t-t')
+2\mean{\hat{a}_s^\dagger(t)\hat{a}_s(t')}{}\\
+2\frac{\mean{\hat{a}_{trig}\hat{a}_s(t)}{}\mean{\hat{a}_{trig}\hat{a}_s(t')}{}}{\mean{\hat{a}_{trig}^\dagger\hat{a}_{trig}}{}}
\end{multline}
where $\hat{a}_{trig}$ corresponds to the operator associated to the idler photon in the mode in which the heralding detection takes place. The first and second terms of Eq. \ref{eq1} are functions of $\abs{t-t'}$ and thus do not come from the conditional operation. The first one corresponds indeed to the vacuum and the second one to the thermal state. In the case of a low pump power, as it is the case in the experiment to limit higher photon number contamination, the contribution from the thermal state is negligible and the autocorrelation function simplifies as:
\begin{equation}
\label{eq2}
\mean{\hat{x}(t)\hat{x}(t')}{}=\delta(t-t')+2\Phi(t)\Phi(t')
\end{equation}
with $\Phi(t)=\frac{\mean{\hat{a}_{trig}\hat{a}_s(t)}{}}{\sqrt{\mean{\hat{a}_{trig}^\dagger\hat{a}_{trig}}{}}}$. Given this form, $\Phi(t)$ is the only eigenfunction with an eigenvalue different from unity. It thus corresponds to the mode of the heralded single-photon. Moreover, if we consider an extremely fast detector, leading to $\hat{a}_{trig}=\int dt \hat{a}_{i}(t)\delta(t)$, the temporal mode becomes $\Phi(t)=\sqrt{\pi\gamma}e^{-|t|\pi \gamma}$ where $\gamma$ is the bandwidth of the OPO cavity. This double-decaying exponential profile was already demonstrated in \cite{Nielsen2007_1,Sasaki} but using different methods. 

Given the experimental bandwidth of the OPO used here ($\gamma=60$ MHz), the theoretical temporal profile is superimposed on Fig. \ref{fig2}(b). Due to a very optimized experimental setup, the results are very close to this simplest theoretical model: the overlap between experimental and theoretical temporal modes is above 99\%. A tiny difference can be observed on the profile. The experimental function are indeed smoother on the top. This could be explained by some jitter on the detection signal but the superconducting single-photon detector used for this experiment is around 50 ps. The main reason comes from the limited bandwidth of the homodyne detection which is larger but not much larger than the one of the OPO cavity. The very good agreement obtained here in the case of a well-controlled system clearly confirms the reliability of the method, providing the optimal temporal mode without prior information. Our method can therefore be applied to more involved cases where the temporal modes is completely unknown or difficult to predict. 

We also applied this practical method to the generation of a coherent state superposition, which includes higher photon numbers. This generation can be heralded by the subtraction of a single-photon from a squeezed vacuum \cite{Dakna, Grangier,Polzik,Furusawa}. For this purpose, we tap out with a beam splitter 5\% of a 3 dB-squeezed vacuum generated by a type-I optical parametric oscillator and, as previously, herald the preparation by a detection event given by the superconducting single-photon detector. In contrast to the previous case, the generated state is not phase invariant but the autocorrelation function is computed with a phase-average of the quadrature measurements. Indeed, the temporal mode does not depend on the quadrature and the phase averaging guarantees to obtain eigenvalues above the vacuum one, a requirement for extracting the modes. The eigenvalues are given in Fig. \ref{fig2}(a) and only one mode with a large eigenvalue appears. The other modes are squeezed vacuum leading to eigenvalues slightly larger than the vacuum state. The associated temporal profile is given in Fig. \ref{fig2}(b) and the Wigner function corresponding to the heralded single-mode state is displayed in Fig. \ref{fig2}(d).  

\begin{figure}[b!]
\centerline{\includegraphics[width=0.95\columnwidth]{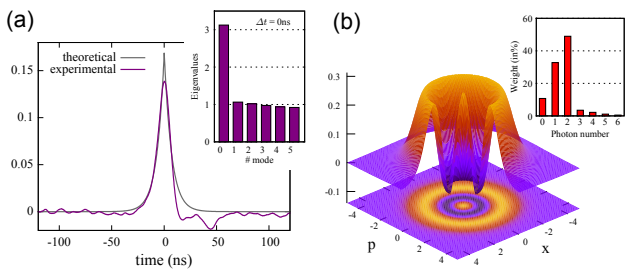}}
\caption{(color online). Two-photon Fock state generation with a delay $\Delta t =0$ between the two heralding events. (a) Eigenvalues of the experimental autocorrelation function (calculated from 10000 heralding events) and optimal temporal mode. (b) Associated Wigner function. The inset shows the diagonal elements of the density matrix, without correction for detection losses.}
\label{fig3}
\end{figure}

As a third illustrative experiment, we consider the more complex case corresponding to the generation of a two-photon Fock state \cite{Ourjoumtsev06,Zavatta08,Bimbard10}. The setup we implemented is the same as for the single-photon generation but the conditioning path is now split into two in order to detect two single-photons as heralding events. In the continuous-wave regime, the two heralding events can occur at different times, $t$ and $t+\Delta t$, and the modal structure strongly depends on this delay.

\begin{figure}[b!]
\centerline{\includegraphics[width=0.95\columnwidth]{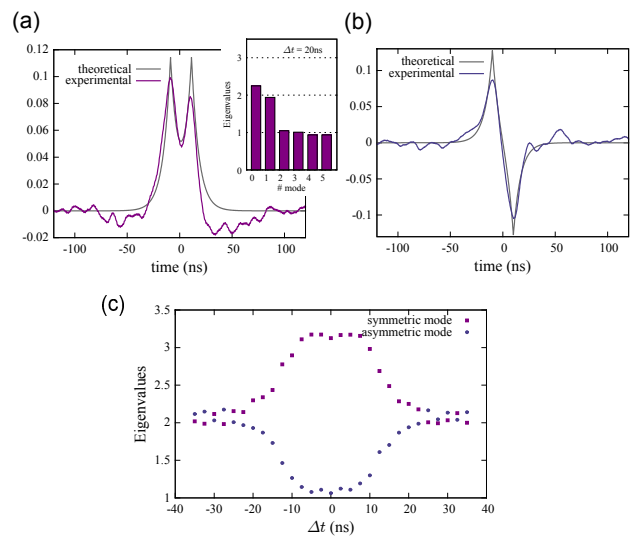}}
\caption{(color online). Two-photon Fock state generation with a delay $\Delta t$ between the two heralding events. In this case, two eigenvalues largely above the vacuum ones appear, as shown in the inset. (a) and (b) provide for $\Delta t =20$ ns the two first eigenmodes, respectively symmetric and antisymmetric. (c) shows the two eigenvalues as a function of the delay. Around $\Delta t =0$, only the symmetric mode has an eigenvalue distinct from the one of a vacuum mode. }
\label{fig4}
\end{figure}

Figure \ref{fig3}(a) gives the results obtained for $\Delta t= 0$. In this case, only one eigenvalue is above the vacuum level and the corresponding temporal profile is similar to the one obtained previously for the single-photon state or the Schr\"odinger cat-like state. Figure \ref{fig3}(b) gives the associated Wigner function and the diagonal elements of the reconstructed density matrix. To the best of our knowledge, this is the first generation of a two-photon Fock state with a two-photon component around 50\% without correction (49$\pm$1\%). By taking into account the detection losses, we infer a value as high as 73$\pm$1\%.

Interestingly, when the time delay $\Delta t$ is different from zero, the situation becomes very different. A second eigenvalue above the vacuum appears: the two-photon Fock state is continuously split between a symmetric and an antisymmetric modes. We give in Fig. \ref{fig4}(a) and Fig. \ref{fig4}(b) the experimental results for $\Delta t =20$ ns. One can notice that the symmetric mode is not perfectly symmetric (different heights). This can be explained by the dark noise being more important on one heralding detector than on the other. This is an example of experimental defaults difficult to take into account into a model. Let us note that the undershoot for positive time (also present in Fig. 3) can be explained by the filtering resulting from the finite bandwidth of the homodyne detection. Figure \ref{fig4}(c) provides the two eigenvalues as a function of the delay $\Delta t$. Around $\Delta t =0$, only the symmetric mode has an eigenvalue distinct from the one of a vacuum mode. Significantly, our method directly provides the exact multimode content of the measured state, and therefore allows us to define an optimal mode.

As before, to compare with a theoretical model, one can use the autocorrelation function defined in Eq. \ref{eq1} and adapt the heralding mode, which now depends on the time separation between the two detection events. In the limit of a low pump power and by considering two fast detection events separated by a delay $\Delta t$, it can be written as:
\begin{multline}
\mean{\hat{x}(t)\hat{x}(t')}{}=\delta(t-t')\\
+2\Phi(t)\Phi(t')+2\Phi(t+\Delta t)\Phi(t'+\Delta t)
\end{multline}
where $\Phi(t)$ has been defined earlier in the single-photon case. Like in Eq. \ref{eq2}, the first term corresponds to the vacuum, and the others to the heralding. The two main eigenfunctions of this kernel are given by:
\begin{equation}
\Psi_{\pm}(t)=\frac{1}{N_{\pm}(\Delta t)}(\Phi(t)\pm \Phi(t+\Delta t)),
\end{equation}
with  $N_{\pm}(\Delta t)$ the normalization factor. All other functions orthogonal to these modes are eigenfunctions for the vacuum. The theoretical profiles are superimposed in Fig. 4.

In conclusion, we have investigated a method to directly access the complete modal content of traveling quantum light states, using raw measurements obtained by homodyning with a continuous-wave local oscillator. This continuous measurement enables a multimode expansion using the autocorrelation function, without making any assumptions on the experimental setup. We provided a detailed study of this method for various quantum state engineering experiments, leading therefore to a very high fidelity state generation. These examples clearly show the efficiency of the procedure to achieve an optimal mode matching taking directly into account all practical aspects. Apart from this crucial aspect, this approach also reveals the single-mode or multimode character of the field states. Let us underline that this method is applicable when the temporal mode is longer than the response time of the detector. Given the current available homodyne detections, with bandwidth up to a few hundreds of MHz, our method can apply to pulses down to the ns regime, which typically includes a large class of experiment, ranging from down conversion in cavity to photon emission by atomic systems. Owing to its simplicity and efficiency, this method will thereby find immediate applications, not only for quantum state engineering but also for optimizing the subsequent use of those states in quantum information processing schemes.\\

\acknowledgements 
This work is supported by the ERA-NET CHIST-ERA (QScale) and the ERC starting grant HybridNet. C. Fabre and J. Laurat are members of the Institut Universitaire de France.

\end{document}